\documentclass[3p,times,twocolumn]{elsarticle}

\usepackage{ecrc}

\volume{00}

\firstpage{1}

\journalname{Nuclear Physics B Proceedings Supplement}

\runauth{L. Yi}

\jid{nppp}

\jnltitlelogo{Nuclear Physics B Proceedings Supplement}

\usepackage{amssymb}

\usepackage{lineno}

\usepackage[figuresright]{rotating}

\usepackage{comment}

\begin{document}

\begin{frontmatter}



\dochead{}

\title{STAR Overview of Hard Probe Observables}


\author{Li Yi (for the STAR Collaboration)}

\address{Physics Department, Yale University, New Haven, CT, USA}

\begin{abstract}
Parton energy loss, quarkonium sequential melting and particle production from electromagnetic interactions are tools to study Quark Gluon Plasma properties. The STAR detector, with large acceptance at mid-rapidity, excellent particle identification and wide transverse momentum coverage, is able to study these probes in details. In Hard Probes 2015, the STAR collaboration reported measurements of reconstructed jets, heavy-flavor physics, di-lepton production and the performance of new detectors in seven presentations and one poster. Given the rich results from STAR, this overview report will focus on a few selected results on jets and $\Upsilon$ measurements in Au+Au collisions at $\sqrt{s_{\rm NN}}=200\,$GeV, $J/\psi$ production in $p$+$p$ collisions at $\sqrt{s_{\rm NN}}=500\,$GeV, and the di-electron spectrum in the low mass region from the  Beam Energy Scan - Phase I.

\end{abstract}


\end{frontmatter}


\section{Introduction}
\label{sec:intro}
High-transvere-momentum ($p_{T}$) partons, produced in the hard scatterings of relativistic heavy-ion collisions, lose energy in and exchange color with the Quark Gluon Plasma (QGP). However, single partons can not be measured experimentally. Early strategies included using high-$p_{T}$ hadrons to study parton energy loss. By comparing high-$p_{T}$ hadrons in heavy-ion and $p$+$p$ collisions, information about parton interactions with the QGP is obtained. The observation of single high-$p_{T}$ hadron suppression in Au+Au collisions matches parton energy loss expectations \cite{Adams:2003kv}. The disappearance of away-side jet peaks in dihadron correlations in Au+Au collisions, but not in $d$+Au collisions, further suggests such high-$p_{T}$ hadron suppression is a hot medium effect \cite{Adams:2003im}. However, the leading hadron does not carry all the information of its parent parton. Full jet reconstruction of hadrons fragmented from the same parton should better represent the kinematics of their parent parton. Therefore, analyses of reconstructed jets are preferred. New results on reconstructed semi-inclusive jets and on dijet imbalance measurements in Au+Au collisions to assess parton energy loss will be reported. 

Besides jet propagation through the QGP as a hard probe, quarkonium production has been used to study the QGP temperature. As a bound state of quark and anti-quark pairs, quarkonia produced in the initial hard scattering can be modified by color screening in the QGP and therefore are sensitive to the QGP temperature. In order to study possible quarkonium modification in the QGP, a reference of quarkonium production in $p$+$p$ collisions is needed. Moreover, data on quarkonium production in $p$+$p$ collisions constrain pQCD calculations \cite{Butenschoen:2011yh,PhysRevLett.113.192301}. $\Upsilon$ ($b\bar{b}$) suppression in Au+Au and U+U collisions will be discussed. $J/\psi$ ($c\bar{c}$) production dependence on event multiplicity at various $p_{T}$ at mid-rapidity in $p$+$p$ collisions will also be discussed to gain insight into small colliding systems. 

Along with top energy running with various collision species for quantitative studies of QGP properties, RHIC has conducted a Beam Energy Scan (BES) program to explore the nuclear matter phase diagram. From 2010 to 2014, RHIC has covered beam energies at $\sqrt{s_{\rm NN}} = $7.7, 11.5, 14.5, 19.6, 27, 39$\,$GeV, together with 62.4, 130, 200$\,$GeV from previous years, which maps the baryon chemical potential in the range of $20<\mu_{B}<420\,$MeV. Results from BES - Phase I have narrowed the region of interest for finding the critical point and the first-order phase boundary to beam energies below 20$\,$GeV \cite{BESWhitePaper}. The di-electron results from BES - Phase I will be discussed to shed light on future measurements. 

\section{Charged Jet Suppression in Au+Au collisions at $\sqrt{s_{\rm NN}}=200\,$GeV}
\label{sec:jet}
Jets link the measured final particles to their parent partons. One of the challenges in jet reconstruction is the enormous underlying event background in heavy-ion collisions, which leads to a considerable number of combinatorial fake jets. To statistically describe the combinatorial jets, a unique mixed event technique has been developed to create events with no correlation between any particles \cite{alex}. Each particle in the mixed event is taken from different events, hence, these mixed events statistically have the same combinatorial jets as those reconstructed within single events. The advantage of using mixed events for background removal is the extension of the jet spectrum to low jet $p_{T}$.  

The semi-inclusive jet measurement is reported with minimum bias (MB) Au+Au collisions at $\sqrt{s_{\rm NN}}=200\,$GeV. Trigger particles are selected within $9 < p_{T} < 30\,$GeV$/c$ to enhance the hard scattering rate for a larger jet signal. The recoil jets are reconstructed at $\Delta\phi=\pi$ with respect to the trigger particle azimuthal angle, and within a $\pi/2$ opening angle. Measuring jets on the recoil side, instead of on the trigger particle side, avoids imposing potential fragmentation biases due to the requirement of a specific high-$p_{T}$ particle inside the jet. Charged particles at mid-rapidity, $|\eta|<1$, measured by the Time Projection Chamber (TPC) are used for jet reconstruction. The anti-$k_{T}$ algorithm from the Fastjet package \cite{fastjet} is applied for jet finding, with a resolution parameter of $R=0.3$. An unfolding procedure is then performed to correct for detector inefficiencies, momentum resolution and residual backgrounds on the measured jet energies.

\begin{figure}[htb]
\includegraphics[width=0.4\textwidth]{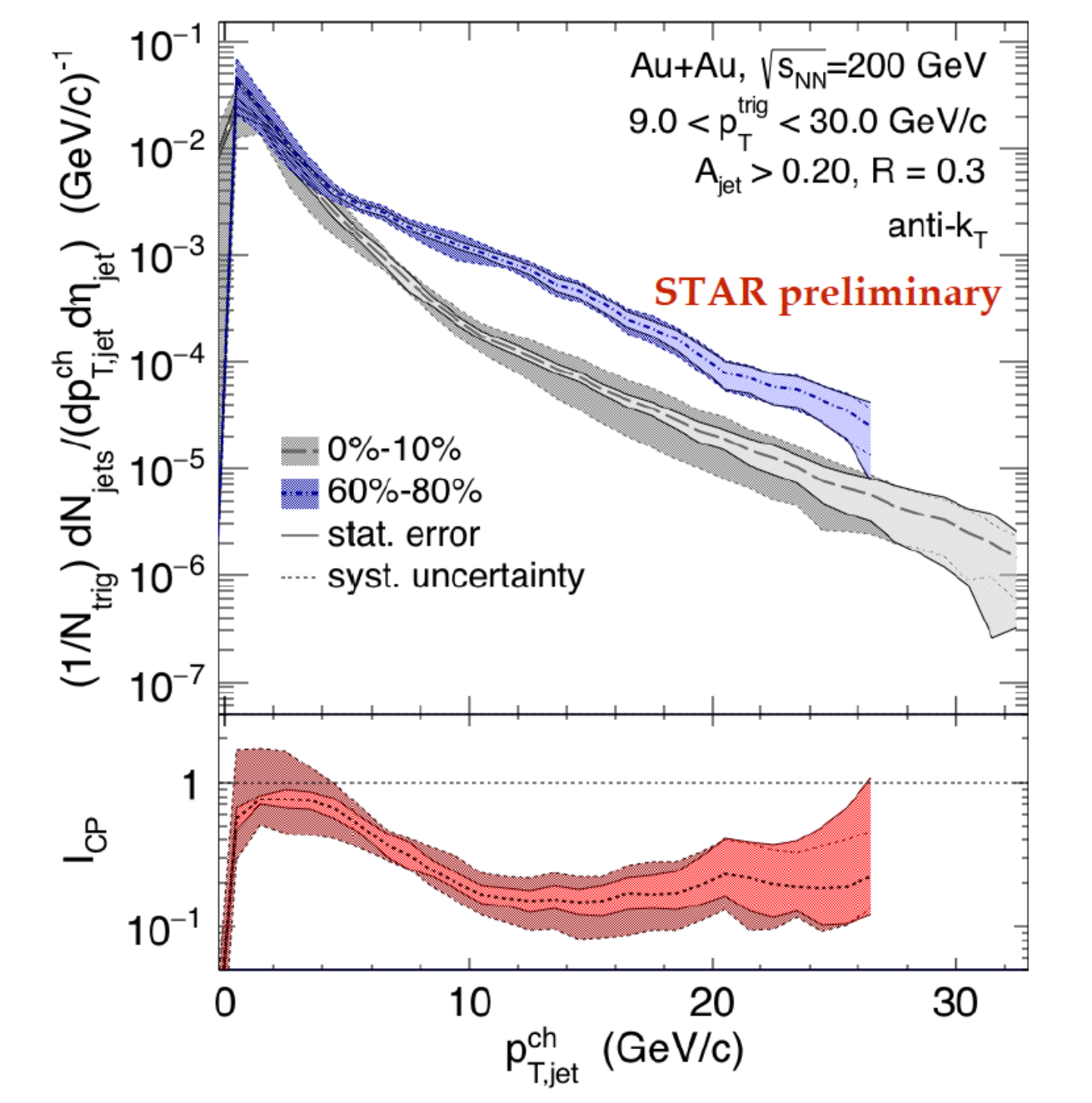}
\caption{Semi-inclusive jet spectra in 0-10\% and 60-80\% central Au+Au 200$\,$GeV collisions are shown in the top panel. The bottom panel shows the ratio between central and peripheral collisions \cite{alex}.}
\label{fig:jet}
\end{figure}

Figure~\ref{fig:jet}, top panel, shows the corrected recoil jet $p_{T}$ spectra for the top 0-10$\%$ centrality class in gray and the 60-80$\%$ centrality class in blue. Dashed curves represent systematic uncertainties and solid curves represent statistical uncertainties. The jet spectrum from the 0-10$\%$ centrality data is lower than the 60-80$\%$ one. $I_{cp}$, shown in the bottom panel of Fig.~\ref{fig:jet}, is the ratio of the jet yields in the 0-10$\%$ and 60-80$\%$ centrality classes. At jet $p_{T} <5\,$GeV$/c$, $I_{cp}$ is close to unity. For jet $p_{T}>10\,$GeV$/c$, $I_{cp}$ drops to $\sim$ 0.2, which indicates significant jet suppression in central Au+Au collisions compared with peripheral collisions. The ALICE collaboration has reported a semi-inclusive charged jet suppression to be around 0.6 for $20<$jet $p_{T}<100\,$GeV$/c$ with $R=0.4$ after recoil jet background subtraction using low-$p_{T}$ trigger particle in Pb+Pb collisions at $\sqrt{s_{\rm NN}}=2.76$ TeV \cite{Adam:2015doa}. This means that the measured jet suppression is stronger at RHIC energies than LHC energies. The jet $p_{T}$ shift, the horizontal shift needed for the peripheral spectra to match the central spectra, however, is similar at RHIC and LHC energies. Therefore, the larger suppression at RHIC may be due to a similar parton energy loss combined with a steeper falling spectrum at RHIC energies than LHC energies. When comparing STAR and ALICE results, the surface bias from different trigger particle $p_{T}$ and collision energies could be different and thus could also impact the measured jet energy loss. 

\section{Dijet Imbalance in Au+Au collisions at $\sqrt{s_{\rm NN}}=200\,$GeV}
\label{sec:dijet}
Dijets are defined in this analysis to be the most energetic leading jet and second most energetic sub-leading jet in a collision. In the absence of a medium, pairs of produced jets have the same $p_{T}$ and are back-to-back with an azimuthal angle difference $\Delta\phi=\pi$ at leading order pQCD level. The medium, however, interacts with and alters jets, i.e. the dijet $p_{T}$ balance is destroyed by medium modifications. Therefore, the comparison of the dijet imbalance 
\begin{equation}
\label{eq:Aj}
A_{j}=\frac{p_{T}^{\rm Leading}-p_{T}^{\rm Sub-Leading}}{p_{T}^{\rm Leading}+p_{T}^{\rm Sub-Leading}}
\end{equation} 
in heavy-ion collisions with that in $p$+$p$ collisions informs us of medium transport properties. 

To enhance the percentage of jet events in the sample, high tower trigger (HT) data are used. HT data are events which have at least one tower of the Barrel Electromagnetic Calorimeter (BEMC) fired with an energy deposit of $E_{T}>5.4\,$GeV. Both tracks in the TPC and tower hits in the BEMC are used to reconstruct the full jet. In order to identify the dijet out of the enormous soft particle background, the first round of anti-$k_{T}$ jet finding with $R=0.4$ is performed with particles of $p_{T}>2\,$GeV$/c$ only. The dijets are required to have an opening angle of $\pi-0.4<|\Delta\phi|<\pi+0.4$. There is no background subtraction since at RHIC energies background are mostly from soft particles with $p_{T}<2\,$GeV$/c$. To select a dijet event, the leading jet is required to have $p_{T}^{\rm Leading}>20\,$GeV$/c$ and the sub-leading jet is required to have $p_{T}^{\rm Sub-Leading}>10\,$GeV$/c$. The 2$\,$GeV$/c$ constituent $p_{T}$ cut followed by a jet $p_{T}$ requirement makes sure that the dijets found are physical jets instead of background fluctuations. After a dijet pair is identified in an event, a more inclusive jet finding with a cut for constituent $p_{T}>0.2\,$GeV$/c$ and $R=0.4$ is performed on these specific jet pair events. This is done by requiring the jets to be within a radial distance of 0.4 in the $\eta-\phi$ plane to the original jets with the constituents cut of $p_{T}>2\,$GeV$/c$. After this selection, the ``median density estimator" \cite{fastjet} is used for background subtraction.

\begin{figure}[htb]
\includegraphics[width=0.4\textwidth]{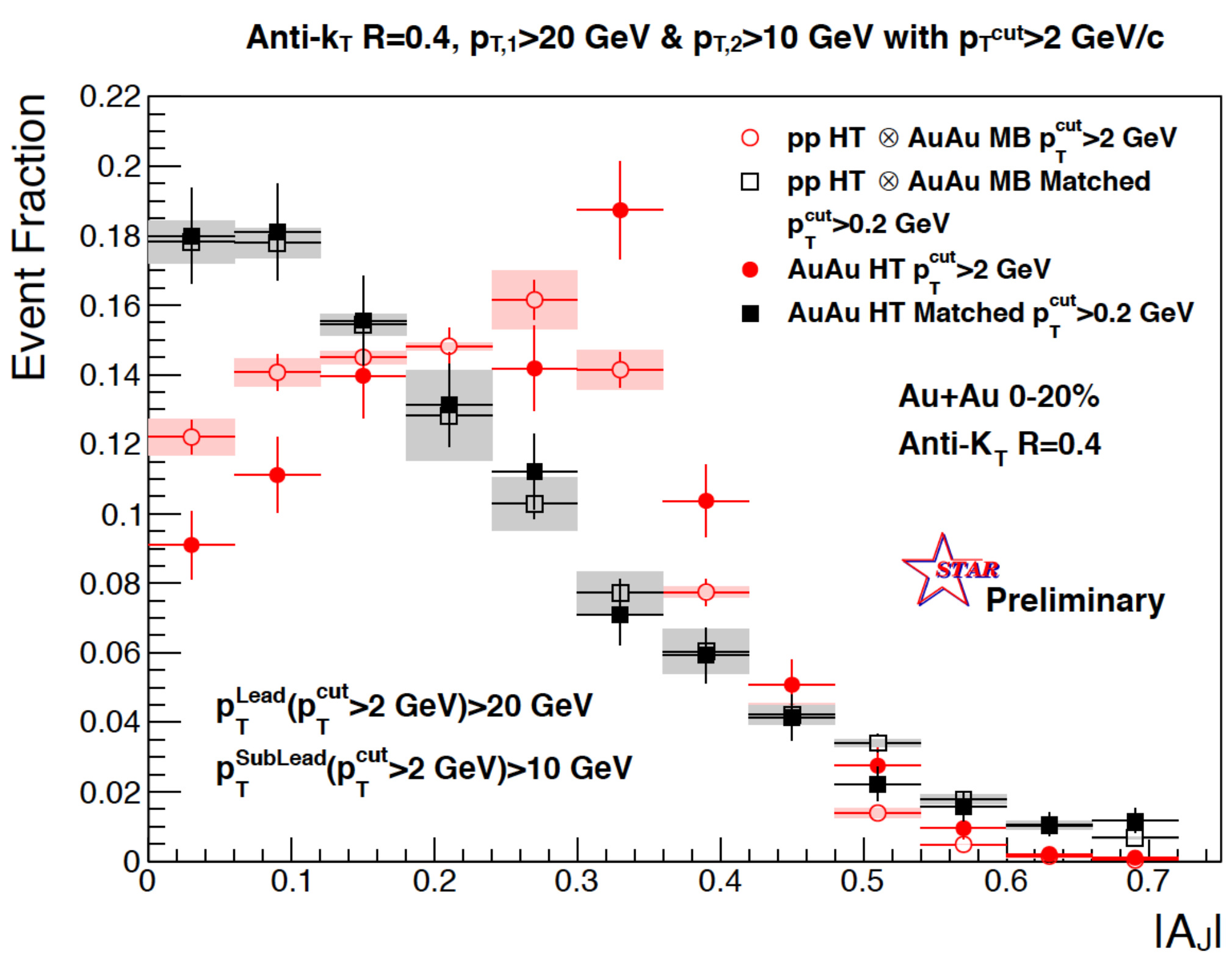}
\caption{Dijet imbalance $|A_{j}|$ distributions for dijet pairs with anti-$k_{T}$ jet finding $R=0.4$. $|A_{j}|$ for jet constituents cuts 0.2$\,$GeV$/c$ and 2$\,$GeV$/c$ are shown for Au+Au collisions and $p$+$p$ embedded into Au+Au collision background (see the text) \cite{kkauder}.}
\label{fig:dijet}
\end{figure}

Figure~\ref{fig:dijet} shows the $|A_{j}|$ of the fully reconstructed jet distributions in red symbols for jets with constituent $p_{T}>2\,$GeV$/c$ and no background subtraction, and in black symbols for constituent $p_{T}>0.2\,$GeV$/c$ and background subtracted. The solid symbols represent $|A_{j}|$ in Au+Au collisions. In order to capture the effect of Au+Au background fluctuations on an unmodified dijet pair with the same kinematic constraints, $p$+$p$ events are embedded into Au+Au collisions. These results are shown as open symbols. Any deviation of the $|A_{j}|$ distribution in Au+Au collisions from the embedded $p$+$p$ data suggests medium modification on the leading and/or sub-leading jets. A $\chi^{2}$-test with the p-value is used to quantify the (dis-)similarity between two distributions. A p-value smaller than $ 0.01$ rejects the hypothesis that the two measured distributions come from the same distribution at the 99\% confidence level. For the jet constituent $p_{T}>2\,$GeV$/c$  case, the p-value is less than $10^{-4}$, which suggests the dijets in Au+Au collisions are different from the dijets in the $p$+$p$ collisions, as expected. However, interestingly, the p-value for the constituent $p_{T}>$ 0.2$\,$GeV$/c$ case is 0.8. This suggests that for the specific dijet set studied here the dijets in Au+Au collisions are similar to the dijets in $p$+$p$ collisions. To interpret this result, one needs to keep in mind that the selection on leading and sub-leading jet $p_{T}$ is expected to pick out special dijet pairs. Theoretical calculations to compare with this result are welcomed. 
When the analyses are repeated with $R=0.2$, the p-value with constituent $p_{T}>$ 0.2$\,$GeV$/c$ reduces to $2\times10^{-4}$, compared with p-value of 0.8 for $R=0.4$ \cite{kkauder}. Dijets can not regain balance with small jet cones which indicates that jets are getting broadened.

\section{$\Upsilon$ suppression in Au+Au and U+U collisions at $\sqrt{s_{\rm NN}}=200\,$GeV}
\label{sec:upsilon}
Quarkonium production in heavy-ion collisions is expected to be sensitive to the QGP energy density. Unlike charmonium production, which has multiple sources, bottomonium, $\Upsilon$, production is a cleaner probe due to negligible contributions from uncorrelated pair recombination or co-mover absorption. Its excited states, which have different binding energies, are predicted to disassociate at different QGP energy densities, thus sequential melting can serve as a QGP thermometer. 

\begin{figure}[htb]
\includegraphics[width=0.4\textwidth]{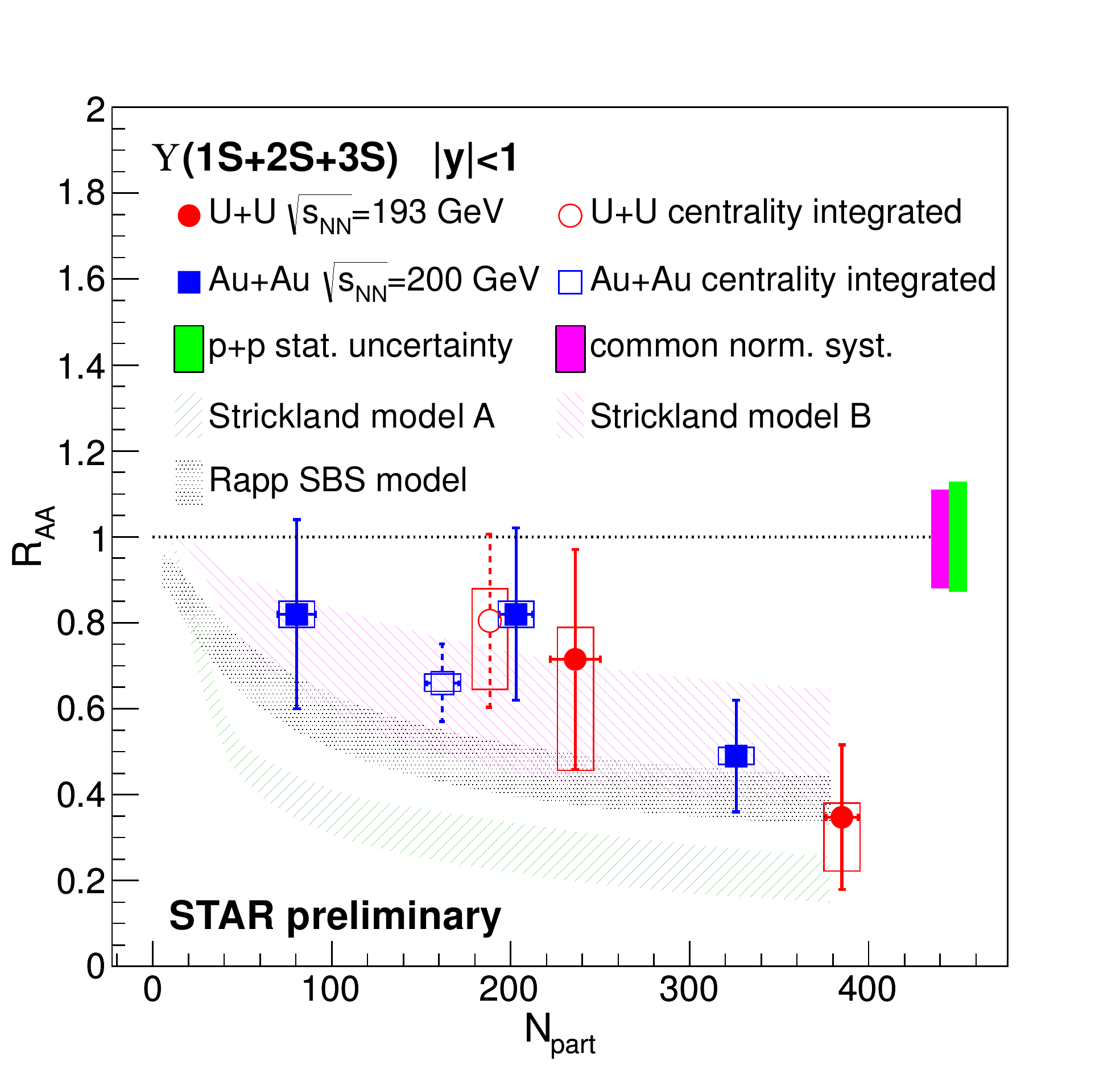}
\caption{$\Upsilon$ nuclear modification factor, $R_{AA}$, vs number of participants, $N_{part}$, in Au+Au 200$\,$GeV and U+U 197$\,$GeV collisions. Several models are also plotted, in which Strickland model B seems to describe data trend better \cite{vertesi}.}
\label{fig:upsilon}
\end{figure}

$\Upsilon$s from the $e^{+} e^{-}$ decay channel are measured with the BEMC detector. Figure~\ref{fig:upsilon} shows the $\Upsilon$ nuclear modification factor, $R_{AA}$, at mid-rapidity as a function of number of participants, $N_{part}$, calculated from the Monte Carlo Glauber Model \cite{Glauber20063}. $R_{AA}$ in Au+Au collisions, at $\sqrt{s_{\rm NN}}=200\,$GeV is shown as blue symbols and $R_{AA}$ in U+U collisions, at $\sqrt{s_{\rm NN}}=197\,$GeV, as red symbols. $R_{AA}$ is close to unity at small $N_{part}$. At larger $N_{part}$ $R_{AA}$ is significantly suppressed. U+U collisions produce higher initial energy densities than Au+Au collisions since the U-nucleus has a larger number of nucleons than a Au-nucleus. The measured U+U data confirm the $\Upsilon$  suppression seen in Au+Au collisions and extend the $R_{AA}$ measurement to higher $N_{part}$. The results of separated $\Upsilon$ states can be found in Ref.~\cite{vertesi}. To test the sequential melting of the different $\Upsilon$ states, more statistics are needed. The newly installed Muon Telescope Detector (MTD) is expected to significantly improve the $\Upsilon$ measurement using the di-muon channel.

\section{$J/\psi$ production event activity dependence in $p$+$p$ collisions at $\sqrt{s_{\rm NN}}=500\,$GeV}
\label{sec:jpsi}
An accurate measurement of $J/\psi$ yields in $p$+$p$ collisions can constrain models of different production mechanisms \cite{Butenschoen:2011yh,PhysRevLett.113.192301} and provide the baseline for heavy-ion collisions. Meanwhile, recent reports from the CMS experiment on $\Upsilon$ yields \cite{Chatrchyan:2013nza} and the ALICE experiment on $D$ meson and $J/\psi$ yields \cite{Adam:2015ota} showed charm and bottom production enhancement in high event activity $p$+$p$ collisions. The event activity was measured by either particle multiplicity or energy deposited in a certain rapidity range.  

\begin{figure}[htb]
\includegraphics[width=0.45\textwidth]{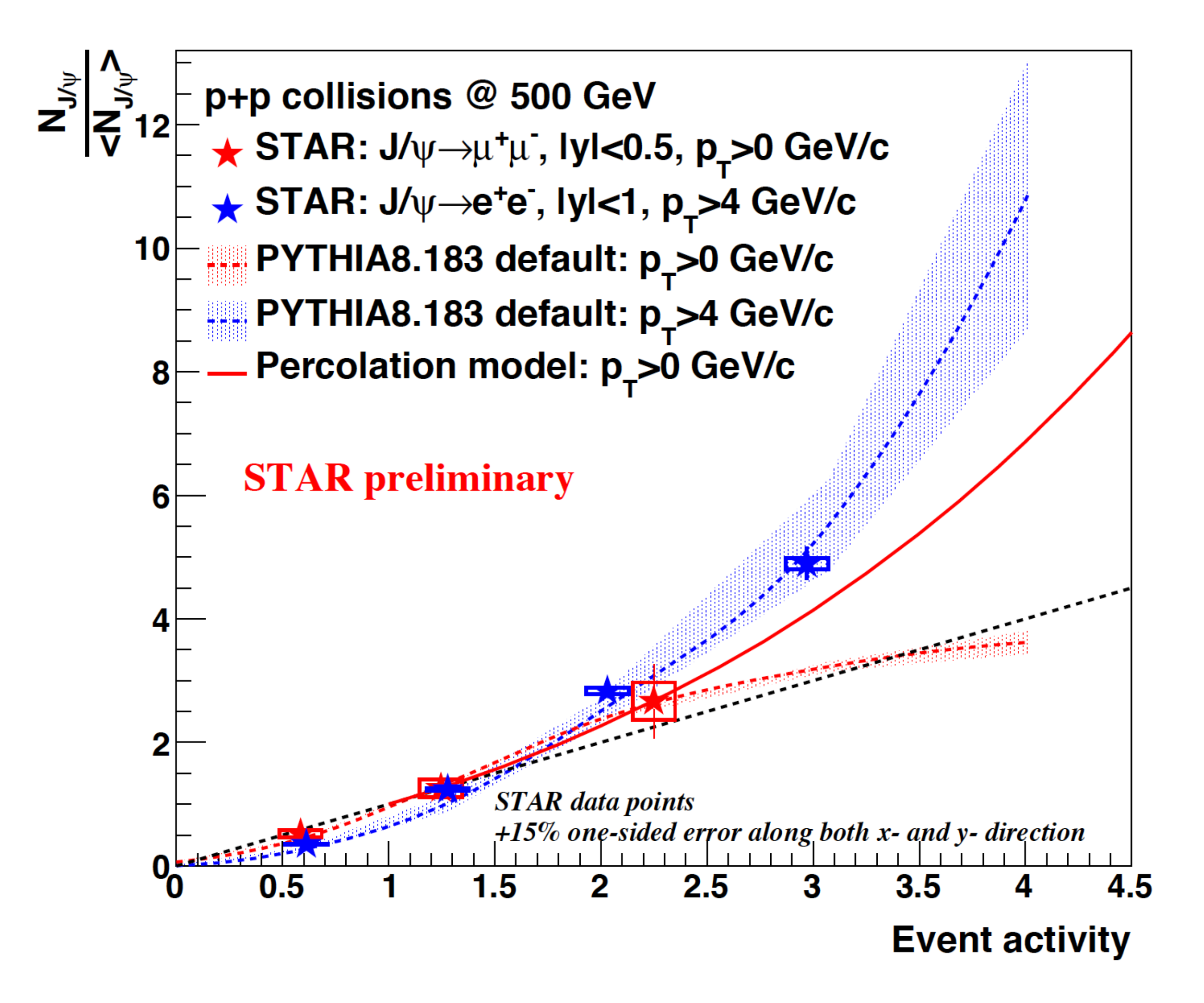}
\caption{The relative $J/\psi$ yields at mid-rapidity vs mid-rapidity relative charged particle multiplicity in 500$\,$GeV $p$+$p$ collisions \cite{rongrong}. }
\label{fig:jpsi}
\end{figure}

Using the MTD for muon identification and the BEMC for electrons together with TPC tracking, STAR measures $J/\psi$ production from both the di-muon and the di-electron channels. Figure~\ref{fig:jpsi} shows relative $J/\psi$ yield as a function of event activity in $p$+$p$ collisions at $\sqrt{s_{\rm NN}}=500\,$GeV. The event activity is characterized by the number of charged particles in $|\eta|<0.9$ relative to that in a MB $p$+$p$ event. The $J/\psi$ production is presented as the relative yield with respect to the one in a MB event. $J/\psi$ production from the di-muon channel with $|\eta|<0.5$ and $p_{T}>0\,$GeV$/c$ are represented as red symbols and $J/\psi$ from the di-electron channel with $|\eta|<1$ and $p_{T}>4\,$GeV$/c$ are the blue symbols. The black dashed line depicts the case of a 1 : 1 ratio of $J/\psi$ production to charged particle multiplicity. The low-$p_{T}$ $J/\psi$ yield in red roughly follows the dashed line for a linear relationship with charged particle multiplicity. The high-$p_{T}$ $J/\psi$ yield, in blue, rises faster than the linear relationship. Enhancement of $J/\psi$ production at high event activity, especially at high-$p_{T}$, implies that hard processes dominate over soft processes in high multiplicity $p$+$p$ collisions relative to low multiplicity collisions. The theoretical calculations shown are from the PYTHIA event generator \cite{Sjostrand:2014zea} and the percolation model \cite{PhysRevC.86.034903}, and are plotted as dashed and solid curves, respectively. For $J/\psi$ with $p_{T}>0\,$GeV$/c$ the two calculations have different trends when the event multiplicity is beyond 2 times the MB multiplicity. More statistics are expected for the high multiplicity range to discriminate models from the year 2015 $p$+$p$ collisions data. 

\section{Di-electron production in Beam Energy Scan}
\label{sec:bes}
Chiral symmetry restoration is one of several signature observables that have been studied to map the nuclear matter phase diagram in the BES program. Low mass di-electron ($e^{+}e^{-}$) production ($<1\,$GeV$/c^{2}$), may characterize in-medium modification of vector mesons, which is a possible link to chiral symmetry restoration. Excess yields in the di-electron invariant mass ($M_{e^{+}e^{-}}$) distributions at $0.2<M_{e^{+}e^{-}}<0.8\,$GeV$/c^{2}$ over the hadronic decay electron cocktails are observed for collision energies from 19.6 to 200$\,$GeV \cite{bingchu}. A model calculation with $\rho$ meson in-medium broadening and chiral symmetry restoration included \cite{rapp} can describe the di-electron spectrum at those energies \cite{bingchu}. Beam energies lower than $20\,$GeV need more data for QGP phase transition searches in BES - Phase II .

\section{Summary and the future}
\label{sec:future}
In summary, larger jet suppression has been measured at RHIC energies than LHC energies for semi-inclusive charged jets, which however corresponds to a similar horizontal $p_{T}$ shift in the jet spectrum. Jet substructures are indicated to become broadened for the hardcore di-jet pairs in Au+Au 200$\,$GeV collisions. $\Upsilon$ suppression in higher energy density U+U collisions confirms and extends the trend seen in Au+Au collisions. The event activity dependence of $J/\psi$ production in $p$+$p$ collisions is observed.

In 2014, STAR underwent two major heavy flavor detector upgrades, the MTD, to identify muons from quarkonium decay, and the HFT to precisely determine low momentum open heavy quark decay vertices for a cleaner signal. The first STAR $J/\psi$ measurement with the MTD is reported in these proceedings \cite{rongrong}. The performance of the HFT for $D^{0}$ signal in $K\pi$ invariant mass distributions can also be found here \cite{qiu}. In 2015-2017, with STAR's enhanced capabilities in heavy flavor measurement, high multiplicity $p$+$p$ and $p$+Au collision will be investigated, and information on the QGP temperature and shear viscosity in Au+Au collisions at 200$\,$GeV are also expected to come.

In 2019-2020, RHIC will have a dedicated program for BES - Phase II with high luminosity for 5-20$\,$GeV Au+Au collisions, capitalizing on the RHIC electron cooling upgrade. Meanwhile, STAR will carry out an inner-TPC (iTPC) upgrade, which will extend the pseudo-rapidity coverage, increase low $p_{T}$ reach in tracking of particles for QGP bulk property study, 
and improve the $dE/dx$ resolution of proton/kaon separations at high $p_{T}$ for jet fragmentation measurements, in addition to the electron identification and acceptance for the low mass di-electron  studies discussed above.
The refined high-precision measurements will enable BES - Phase II to boost our understanding of the nuclear matter phase diagram.




\nocite{*}
\bibliographystyle{elsarticle-num}
\bibliography{Yi_LiYL}

\begin{thebibliography}{10}
\expandafter\ifx\csname url\endcsname\relax
  \def\url#1{\texttt{#1}}\fi
\expandafter\ifx\csname urlprefix\endcsname\relax\def\urlprefix{URL }\fi
\expandafter\ifx\csname href\endcsname\relax
  \def\href#1#2{#2} \def\path#1{#1}\fi

\bibitem{Adams:2003kv}
J.~Adams, {\it et al.}~(STAR~Collaboration), {Transverse momentum and collision
  energy dependence of high $p_{T}$ hadron suppression in Au+Au collisions at
  ultrarelativistic energies}, Phys. Rev. Lett. 91 (2003) 172302.
\newblock \href {http://arxiv.org/abs/nucl-ex/0305015}
  {\path{arXiv:nucl-ex/0305015}}, \href
  {http://dx.doi.org/10.1103/PhysRevLett.91.172302}
  {\path{doi:10.1103/PhysRevLett.91.172302}}.

\bibitem{Adams:2003im}
J.~Adams, {\it et al.}~(STAR~Collaboration), {Evidence from d + Au measurements
  for final state suppression of high $p_{T}$ hadrons in Au+Au collisions at
  RHIC}, Phys. Rev. Lett. 91 (2003) 072304.
\newblock \href {http://arxiv.org/abs/nucl-ex/0306024}
  {\path{arXiv:nucl-ex/0306024}}, \href
  {http://dx.doi.org/10.1103/PhysRevLett.91.072304}
  {\path{doi:10.1103/PhysRevLett.91.072304}}.

\bibitem{Butenschoen:2011yh}
M.~Butenschoen, B.~A. Kniehl, {World data of $J/\psi$ production consolidate
  NRQCD factorization at NLO}, Phys. Rev. D84 (2011) 051501.
\newblock \href {http://arxiv.org/abs/1105.0820} {\path{arXiv:1105.0820}},
  \href {http://dx.doi.org/10.1103/PhysRevD.84.051501}
  {\path{doi:10.1103/PhysRevD.84.051501}}.

\bibitem{PhysRevLett.113.192301}
Y.-Q. Ma, R.~Venugopalan, Comprehensive description of $\ensuremath{J/\psi}$
  production in proton-proton collisions at collider energies, Phys. Rev. Lett.
  113 (2014) 192301.
\newblock \href {http://dx.doi.org/10.1103/PhysRevLett.113.192301}
  {\path{doi:10.1103/PhysRevLett.113.192301}}.

\bibitem{BESWhitePaper}
{STAR collaboration},
  \href{https://drupal.star.bnl.gov/STAR/starnotes/public/sn0598}{{Beam Energy
  Scan II White Paper: Studying the Phase Diagram of QCD Matter at RHIC}}.
\newline\urlprefix\url{https://drupal.star.bnl.gov/STAR/starnotes/public/sn0598}

\bibitem{alex}
A.~Schmah, these proceedings.

\bibitem{fastjet}
M.~Cacciari, G.~P. Salam, G.~Soyez, {FastJet User Manual}, Eur. Phys. J. C72
  (2012) 1896.
\newblock \href {http://arxiv.org/abs/1111.6097} {\path{arXiv:1111.6097}},
  \href {http://dx.doi.org/10.1140/epjc/s10052-012-1896-2}
  {\path{doi:10.1140/epjc/s10052-012-1896-2}}.

\bibitem{Adam:2015doa}
J.~Adam, {\it et al.}~(ALICE~Collaboration), {Measurement of jet quenching with
  semi-inclusive hadron-jet distributions in central Pb-Pb collisions at
  ${\sqrt{\bf{s}_{\mathrm {\bf{NN}}}}}$ = 2.76 TeV}\href
  {http://arxiv.org/abs/1506.03984} {\path{arXiv:1506.03984}}.

\bibitem{kkauder}
K.~Kauder, these proceedings.

\bibitem{vertesi}
R.~Vertesi, these proceedings.

\bibitem{Glauber20063}
R.~J. Glauber, Quantum optics and heavy ion physics, Nuclear Physics A 774
  (2006) 3 -- 13, {QUARK} {MATTER} 2005 Proceedings of the 18th International
  Conference on Ultra-Relativistic Nucleus--Nucleus Collisions.
\newblock \href
  {http://dx.doi.org/http://dx.doi.org/10.1016/j.nuclphysa.2006.06.009}
  {\path{doi:http://dx.doi.org/10.1016/j.nuclphysa.2006.06.009}}.

\bibitem{Chatrchyan:2013nza}
S.~Chatrchyan, {\it et al.}~(CMS~Collaboration), {Event activity dependence of
  $\Upsilon$(nS) production in $\sqrt{s_{NN}}$=5.02 TeV pPb and $\sqrt{s}$=2.76
  TeV pp collisions}, JHEP 04 (2014) 103.
\newblock \href {http://arxiv.org/abs/1312.6300} {\path{arXiv:1312.6300}},
  \href {http://dx.doi.org/10.1007/JHEP04(2014)103}
  {\path{doi:10.1007/JHEP04(2014)103}}.

\bibitem{Adam:2015ota}
J.~Adam, {\it et al.}~(ALICE~Collaboration), {Measurement of charm and beauty
  production at central rapidity versus charged-particle multiplicity in
  proton-proton collisions at $\mathbf{\sqrt{{\textit s}}}=7$ TeV}\href
  {http://arxiv.org/abs/1505.00664} {\path{arXiv:1505.00664}}.

\bibitem{rongrong}
R.~Ma, these proceedings.

\bibitem{Sjostrand:2014zea}
T.~Sj{\"o}strand, S.~Ask, J.~R. Christiansen, R.~Corke, N.~Desai, P.~Ilten,
  S.~Mrenna, S.~Prestel, C.~O. Rasmussen, P.~Z. Skands, {An Introduction to
  PYTHIA 8.2}, Comput. Phys. Commun. 191 (2015) 159--177.
\newblock \href {http://arxiv.org/abs/1410.3012} {\path{arXiv:1410.3012}},
  \href {http://dx.doi.org/10.1016/j.cpc.2015.01.024}
  {\path{doi:10.1016/j.cpc.2015.01.024}}.

\bibitem{PhysRevC.86.034903}
E.~G. Ferreiro, C.~Pajares, High multiplicity $pp$ events and
  $\ensuremath{J/\psi}$ production at energies available at the {CERN} large
  hadron collider, Phys. Rev. C 86 (2012) 034903.
\newblock \href {http://dx.doi.org/10.1103/PhysRevC.86.034903}
  {\path{doi:10.1103/PhysRevC.86.034903}}.

\bibitem{bingchu}
B.~Huang, these proceedings.

\bibitem{rapp}
R.~Rapp, priviate communication.

\bibitem{qiu}
H.~Qiu, these proceedings.

\bibitem{DEramo:2012jh}
F.~D'Eramo, M.~Lekaveckas, H.~Liu, K.~Rajagopal, {Momentum Broadening in Weakly
  Coupled Quark-Gluon Plasma (with a view to finding the quasiparticles within
  liquid quark-gluon plasma)}, JHEP 05 (2013) 031.
\newblock \href {http://arxiv.org/abs/1211.1922} {\path{arXiv:1211.1922}},
  \href {http://dx.doi.org/10.1007/JHEP05(2013)031}
  {\path{doi:10.1007/JHEP05(2013)031}}.

\end{thebibliography}







\end{document}